\documentclass[aps,prb,twocolumn,showpacs,superscriptaddress,groupedaddress]{revtex4-1}  

\usepackage{graphicx}  
\usepackage{dcolumn}   
\usepackage{bm}        
\usepackage{amssymb}   
\usepackage{braket}
\usepackage{amsmath}   
\usepackage{color}
\usepackage{ulem}

\begin{document}

\newcommand{\Kvec}{\bm{K}}
\newcommand{\kvec}{\bm{k}}
\newcommand{\kvecSC}{\bm{k}_{\rm SC}} 
\newcommand{\Gvec}{\bm{G}}
\newcommand{\lvec}{\bm{l}}
\newcommand{\tvec}{\bm{t}}
\newcommand{\Lvec}{\bm{L}}
\newcommand{\muvec}{\bm{\mu}}
\newcommand{\rvec}{\bm{r}}




\title{Band-unfolding approach to Moir\`{e}-induced band-gap opening and Fermi-level-velocity reduction in twisted bilayer graphene}

\author{Hirofumi Nishi}
\author{Yu-ichiro Matsushita}
\email{matsushita@ap.t.u-tokyo.ac.jp}
\author{Atsushi Oshiyama}

\affiliation{
Department of Applied Physics, The University of Tokyo, Hongo, Tokyo 113-8656, Japan
 }

\date{\today}

\begin{abstract}
We report on the energy spectrum of electrons in twisted bilayer graphene (tBLG) obtained by the band-unfolding method in the tight-binding model. We find the band-gap opening at particular points in the reciprocal space, that elucidates the drastic reduction of the Fermi-level velocity with the tiny twisted angles in tBLGs. We find that Moir\`e pattern caused by the twist of the two graphene layers generates interactions among Dirac cones, otherwise absent, and the resultant cone-cone interactions peculiar to each point in the reciprocal space causes the energy gap and thus reduced the Fermi-level velocity. 
\end{abstract}

\maketitle

\section{Introduction}

Graphene is a carbon sheet in which $sp^2$ hybridized electrons ($\sigma$ electrons) form a honeycomb structure and the remaining $\pi$ electrons follow the massless Dirac equation (Weyl equation). The energy bands show the linear dispersion (Dirac cone) near the Fermi level ($E_{\rm F}$) around particular symmetry points, $K$ and $K^{\prime}$, in Brillouin zone (BZ) \cite{SW}. This causes unusual properties such as the anomalous quantum Hall effect \cite{QHE1,QHE2} and the unexpected magnetic ordering \cite{louie,okada}. Further its in-plane high mobility and the structural robustness place graphene as an emerging material in the post-scaling semiconductor technology\cite{device}. 
\par
Two sheets of graphene are stacked via van der Waals interaction to form bilayer graphene (BLG). BLG is in fact produced by the exfoliation of graphite \cite{Exfoliation1,Exfoliation2} or by the heat treatment of SiC \cite{SiC1,SiC2,SiC3}. Dirac cones are modified in BLG depending on the way of the stacking: In the AA (on top) stacking the interlayer interaction induces the bonding (antibonding) $\pi$ orbitals, generating the lower (upper) Dirac cones, whereas the non-bonding $\pi$ orbitals peculiar to the AB (Bernal) stacking result in the parabolic bands at $E_{\rm F}$ \cite{ABband}. 
\par
The stacking is not limited to the AA and the AB owing to the relatively weak interlayer interaction. The two graphene layers are generally twisted to each other in their basal planes. The layers with a tiny twisted angle $\theta$ generate a Moir\`e pattern with its period $L=\sqrt{3}d /(2 \sin \frac{\theta}{2})$ where $d$ is the C-C bond length \cite{ Exfoliation1,Exfoliation2,SiC1,SiC2,SiC3, STM1993,STM2005,ARPES2012,hicks,Uchida}. When the twisted angle ($ 0^{\circ} \le \theta \le 60^{\circ} $) is sufficiently far from either $0^{\circ}$ (AA stacking) or $60^{\circ}$ (AB stacking), the two layers are practically decoupled \cite{Exfoliation2,SiC1,SiC2,ARPES2009,latil,Pankratov2008}. However, when the two layers are slightly twisted from either the AA or the AB stackings by less than $5^{\circ}$ \cite{Uchida}, the drastic reduction of the Fermi velocity $v_{\rm F}$ is experimentally observed \cite{Exfoliation1,SiC3} and theoretically predicted \cite{Uchida,Neto2007,Pankratov2010,Magaud2010,Barticevic2010,MacDonald2011,Magaud2012,Neto2012}. Some magic twisted angles around and below 1$^{\circ}$ at which the Fermi velocity vanishes are also predicted theoretically \cite{Uchida,MacDonald2011,Magaud2012}. 
\par
The origin of this mysterious modification of the Fermi velocity in the twisted bilayer graphene (tBLG) has been discussed: The continuum theory shows that the interactions between the Dirac cones located around different $K$ points causes this reduction \cite{MacDonald2011}; the tight binding calculation \cite{Magaud2010} and the density-functional calculation \cite{Uchida} show that the orbital at $E_{\rm F}$ is localized in the AA stacking region, thus $v_{\rm F}$ being reduced. However, the relation between those two arguments is unclear yet. Furthermore, the reduction is obviously caused by the Moir\`e pattern but the underlying physics of Moir\`e causing the reduction is still a mystery. It is thus highly demanded to connect the effective continuum theory with the atomistic electronic-structure theory and then clarify the microscopic origin of the $v_{\rm F}$ reduction and moreover the Moir\`e-induced modification of the electronic structure in tBLG. 
\par
Atomistic calculations for tBLG with the small twist angle have been formidable. The periodicity of the Moir\`e pattern is generally incommensurate with the periodicity of each graphene layer. When we constrain ourselves to the particular twist angles which generate the commensurate tBLG, we may play with the supercell model. Even in the case, the system considered is gigantic composed of tens of thousands of atoms, as is evident from the relation between the Moir\`e periodicity $L$ and the twist angle $\theta$ above. However, recent approach of {\it computics} \cite{computics,hasegawa} allows us to perform such large-scale density-functional calculations \cite{Uchida}. Yet the analyses to clarify the microscopic origin of the Moir\`e induced physics is still challenging since the unit cell of the commensurate tBLG contains a huge number of atoms and the corresponding BZ is tiny, that hinder such analyses.

A promising theoretical scheme for such analyses is the band-unfolding scheme \cite{Ku2010,Zunger2010,Zunger2012,Sawatzkey,Ozaki} which have been originally used to analyze electronic structures of alloys and of defects in solids. In the scheme, the supercell model is introduced to describe the system and the calculated ``energy bands" in the supercell BZ are unfolded to the original BZ of the reference system: e.g., for the defect in a solid, the supercell containing the defect and host atoms is introduced and the calculated energy bands of the supercell are unfolded to BZ of the host solid. In the unfolding procedure, we calculate the spectral function expressed by the eigen-function of the reference system. Hence the obtained unfolding bands in the original BZ with their particular spectral weights represent the contribution from the eigenstates of the reference system to the energy spectrum of the target supercell system.

In this paper, we apply the band-unfolding scheme within the tight-binding model to commensurate tBLG with various twisted angles $\theta$. For each tBLG, the supercell energy bands are unfolded to the original BZ of one of the two constituting graphene monolayers. The obtained unfolded bands unequivocally reveal that the original Dirac cones at two graphene layers interact with each other at some particular points in BZ, open a energy gap there near $E_{\rm F}$, and eventually cause the reduction of $v_{\rm F}$.

The organization of this paper is as follows. We explain pertinent features of our band-unfolding scheme in the TB model in section \ref{method}. The obtained unfolded bands are presented and the mechanism of the modification of the electronic structures by the Moir\`e pattern is analyzed in section \ref{results}. Section \ref{sum} summarizes our findings.

\section{Band unfolding method}\label{method}

We perform band-structure calculations for tBLGs with the twisted angles of 1.47$^{\circ}$, 1.89$^{\circ}$, 2.88$^{\circ}$, 3.48$^{\circ}$, 3.89$^{\circ}$, 4.41$^{\circ}, $5.09$^{\circ}$, 6.01$^{\circ}$, 9.34$^{\circ}$, 13.17$^{\circ}$ and 32.20$^{\circ}$ which correspond to the supercells consisting of 6076, 3676, 1588, 1048, 868, 676, 508, 364, 148, 76 and 52 atoms, respectively. 
Then we unfold the obtained bands to the primitive BZ of a monolayer graphene. In the band-structure calculations, we adopt a tight-binding (TB) model \cite{Magaud2010,Magaud2012} for carbon $\pi$ orbitals since the electronic structure near the Fermi level is satisfactorily described by the linear combination of the $\pi$ orbitals. 
\par
The band-unfolding scheme relies on the one-particle Green's function of the supercell (SC) system which is given by 
\begin{equation}
  \hat{G}(z) = \sum_{\kvecSC I} \frac{ \Ket{\Psi_{I}^{\rm{SC}} (\kvecSC)} 
\Bra{\Psi_{I}^{\rm{SC}} (\kvecSC)} } {z-\epsilon_{\kvecSC I}}  , 
\label{Gfunc}
\end{equation}
where $| \Psi_{I}^{\rm{SC}} (\kvecSC) \rangle$ is the eigenstate of the supercell system labeled by the band index $I$ and the wave vector $\kvecSC$ in the supercell BZ, and $\epsilon_{\kvecSC I} $ is its eigenvalue. 
The supercell contains primitive cells (PCs) of the reference system, i.e., each of the two monolayers of graphene in the present case. Then the wave vector $\kvec$ in the BZ of the reference system (PBZ hereafter) is folded to the supercell BZ (SBZ hereafter) with the reciprocal lattice vector $\Gvec$ of the SC as 
\begin{equation}
\kvecSC = \kvec - \Gvec  \ .
\label{foldunfold}
\end{equation}
It is noteworthy that a wave vector $\kvecSC$ in the SBZ is unfolded to inequivalent several wavevectors $\kvec$ in the PBZ. We next introduce the spectral function as  
\begin{equation}
A (\kvec, \epsilon) = \sum_{i} \langle \psi_i^{\rm PC} (\kvec) |
- \frac{1}{\pi} \mathrm{Im} \hat{G}(\epsilon + \mathrm{i}0^+) 
| \psi_i^{\rm PC} (\kvec) \rangle  \ ,
\label{spec1}
\end{equation}
with $0^+$ being a positive infinitesimal. Here $\psi_i^{\rm PC} (\kvec) $ is the eigenfunction of the reference system (the PC system) labelled by the wavevector $\kvec$ and the band index $i$. Hence this spectral function is a measure of the contributions from the PC eigenstates with the wave vector $\kvec$ to the SC eigenstates. When a PC eigenstate $\psi_i^{\rm PC} (\kvec) $ with the eigenvalue of $\epsilon_{\kvec i}$ is still a good quantum state in the SC system, the spectral function $A (\kvec, \epsilon)$ becomes large around $\epsilon = \epsilon_{\kvec i}$, whereas, when it is an irrelevant quantum state, $A (\kvec, \epsilon)$ becomes small or even vanishing at the energy point. In this sense, this spectral function is an energy spectrum of the SC system unfolded to the PCB: i.e., the unfolded band. The spectral function is written as 
\begin{equation}
A (\kvec, \epsilon) = \sum_{\kvecSC I} P_{\kvecSC I} (\kvec) 
\delta (\epsilon - \epsilon_{\kvecSC I})  \ ,
\label{spec2}
\end{equation}
in terms of the spectral weight, 
\begin{equation}
P_{\kvecSC I} (\kvec) = \sum_{i}|
\Braket{\psi_{i}^{\rm{PC}} (\kvec) | \Psi_{I}^{\rm{SC}} (\kvecSC) } 
|^2  \ .
\label{specweight}
\end{equation}

In the scheme of the linear combination of atomic orbitals (LCAO), the eigenfunction $\psi_i^{\rm PC} (\kvec) $ of the PC is expressed in terms of the $\alpha$-th atomic orbital $\varphi_{\alpha \muvec} (\rvec) $ located at the site $\muvec$ in the PC: i.e., 
\begin{equation}
\psi_i^{\rm PC} (\kvec) = \sum_{\alpha \muvec} u_{i \kvec}^{\alpha \muvec} 
\phi_{\alpha \muvec} (\kvec) \ , 
\label{LCAOPC}
\end{equation}
where 
\begin{equation}
\phi_{\alpha \muvec} (\kvec) = \frac{1}{\sqrt{n_c}} 
\sum_{\lvec} e^{{\rm i} \kvec \cdot \lvec} 
\varphi_{\alpha \muvec} (\rvec - \lvec ) \ .
\label{LCAObase}
\end{equation}
Here $\lvec$ and $n_c$ are the lattice vector and the number of the primitive cells. The coefficient $ u_{i \kvec}^{\alpha \muvec}$ is determined by solving the secular equation. Similarly for the eigenfunction $\Psi_{I}^{\rm{SC}} (\kvecSC)$ of the SC system is given by 
\begin{equation}
\Psi_I^{\rm SC} (\kvecSC) = \sum_{\alpha \muvec} \sum_{\tvec} 
U_{I \kvecSC}^{\alpha \muvec \tvec}  \Phi_{\alpha \muvec \tvec} (\kvecSC) \ , 
\label{LCAOSC}
\end{equation}
where 
\begin{equation}
\Phi_{\alpha \muvec \tvec} (\kvecSC) = \frac{1}{\sqrt{N_c}} 
\sum_{\Lvec} e^{{\rm i} \kvecSC \cdot \Lvec } 
\varphi_{\alpha \muvec} (\rvec - \Lvec - \tvec ) \ .
\label{LCAObase2}
\end{equation}
Here $\Lvec$ and $N_c$ are the SC lattice vector and the number of the supercells. The $\tvec$ is the PC lattice vector in a single SC, and $U_{I \kvecSC}^{\alpha \muvec \tvec}$ is to be determined by solving the corresponding secular equation. 
Then the spectral weight [Eq. (\ref{specweight})] is expressed in terms of $U_{I \kvecSC}^{\alpha \muvec \tvec}$ and $ u_{i \kvec}^{\alpha \muvec}$ along with overlap integrals. The summation over the band index $i$ is carried out and becomes the matrix of the overlap integrals. When we assume such overlap matrix as an identity matrix (i.e., the tight binding approximation), we obtain
\begin{equation}
\label{eq:TB}
P_{\kvecSC I} (\kvec)  = \frac{N_c}{n_c} \sum_{\alpha \muvec} \sum_{\tvec\tvec^{\prime}} 
\mathrm{e}^{{\rm i} \kvec \cdot (\tvec - \tvec^{\prime})} 
{U_{I\kvecSC}^{\alpha \muvec \tvec}}^{\ast}  {U_{I\kvecSC}^{\alpha\muvec \tvec^{\prime}}}  \ .
\end{equation}

\section{Results and discussions}\label{results}

In this paper we have studied the electronic structures of tBLGs with changing the twisted angle $\theta$ in a range from 1.47$^{\circ}$ to 32.2$^{\circ}$ by using the tight-binding-unfolding method. Those tBLGs are commensurate with each of the graphene monolayers. The periodicity of a tBLG is defined by the two lattice vectors $\bm A_1 = N\bm a_1 +M\bm a_2$ and $\bm A_2=¿M\bm a_1+(N+M)\bm a_2$ where $\bm a_1$ and $\bm a_2$ are the primitive lattice vectors of a monolayer graphene and $N$ and $M$ are integers \cite{Uchida}. The twisted angle is written as
\begin{equation}
  \cos\theta = \frac{N^2+4NM+M^2}{2(N^2+NM+M^2)} \ .
\end{equation}
When $M=N+1$, the $K$ and $K^{\prime}$ points of the PBZs of both the two graphene layers twisted to each other are folded on either $K$ or $K^{\prime}$ of the SBZ. We label symmetry points of the SBZ as, for instance, $K_{\rm SC}$ and $K^{\prime}_{\rm SC}$ hereafter. We present the results for the case of $M=N+1$ in this paper except for the twisted angle $\theta=32.20^{\circ}$ corresponding to $(M,N)=(3,1)$.

\begin{figure}
\includegraphics[bb=0 0 670 519,width=1.0\linewidth]{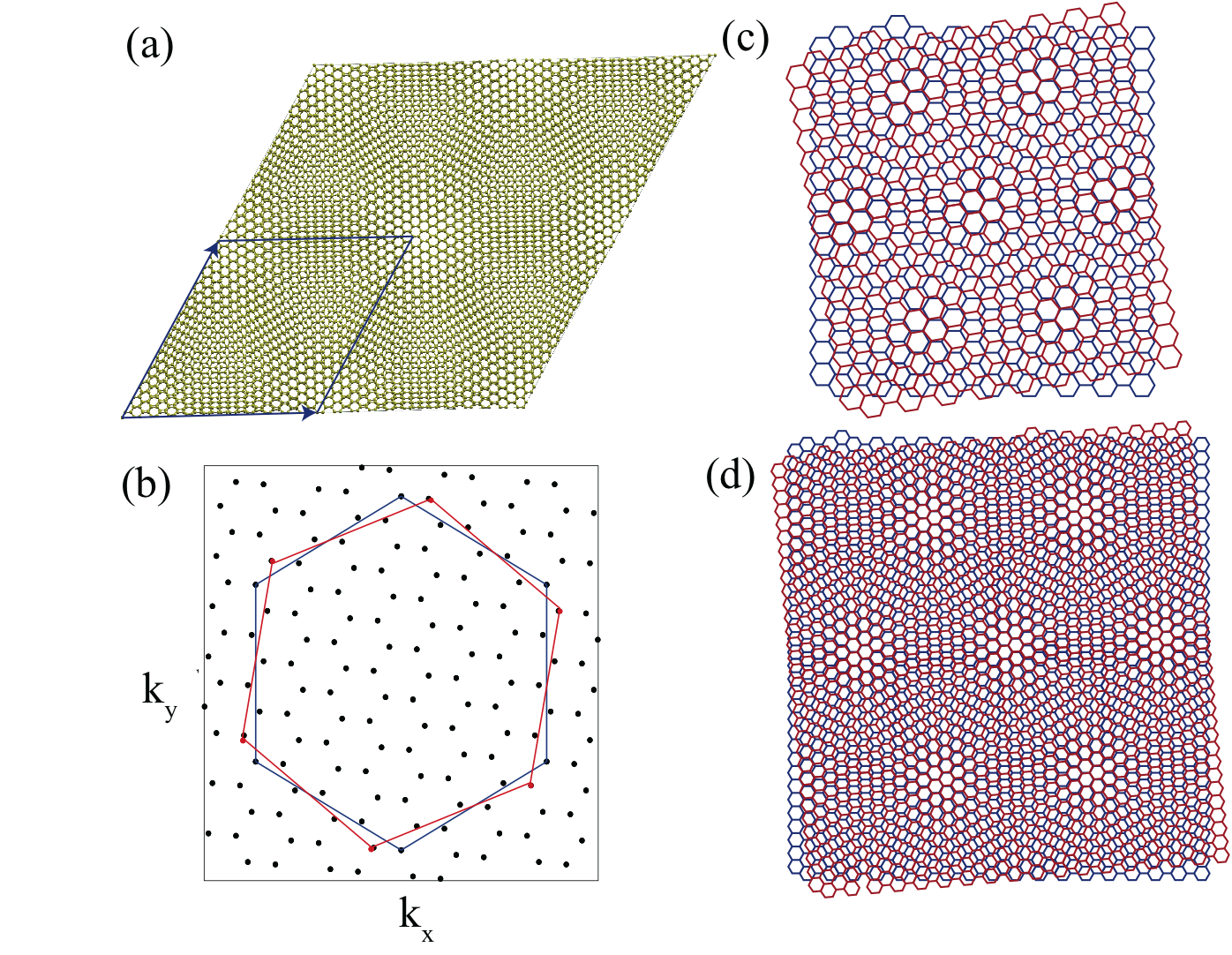}
\caption{(Color online) Atomic structure and BZs of tBLG. (a) Atomic structure of tBLG with $\theta=2.88^{\circ}$ with blue arrows being the lattice vectors of the SC. (b) BZ of tBLG with $\theta=9.43^{\circ}$. Black dots represent either $K_{\rm SC}$ or $K_{\rm SC}^{\prime}$ point of SBZ, and the red and blue lines depict the PBZs of the two graphene monolayers twisted to each other. The BZs in the extended scheme for $\theta=9.43^{\circ}$ (c) and for $\theta=5.09^{\circ}$ (d) are shown.}
\label{fig:atom-bz}
\end{figure}

In Fig. \ref{fig:atom-bz}(a), we show the atomic structure of tBLG with $\theta=2.88^{\circ}$. Moir\`e pattern which is a manifestation of the interference of two slightly different periodicities caused by the twisting is clearly observed. Figure \ref{fig:atom-bz} (b) shows BZ of tBLG with $\theta = 9.43^{\circ}$ corresponding to $N=3$ in which black dots represent either $K_{\rm SC}$ or $K^{\prime}_{\rm SC}$ point the apexes of SBZ and the red and blue lines depict the PBZs of the two graphene monolayers twisted to each other (we call them red and blue PBZ hereafter). The PBZ of each monolayer is twisted to each other by the same angle as in real space. It is noteworthy that the Moir\`e pattern emerges also in the BZ and the pattern becomes prominent when the twisted angle becomes small [Figs.~\ref{fig:atom-bz} (c) and (d)]. 
The preceding work \cite{MacDonald2011} reported that Fermi velocity reduction was reproduced by introducing interaction just between two adjacent Dirac cones from different layers. 
We here point out that the distance between two particular Dirac cones is inappropriate to describe the amount of the interaction since the distance differ from place to place in the extended scheme as in Figs.~\ref{fig:atom-bz} (c) and (d). 

\begin{figure}
\includegraphics[bb = 0 0 527 287,width=1.0\linewidth]{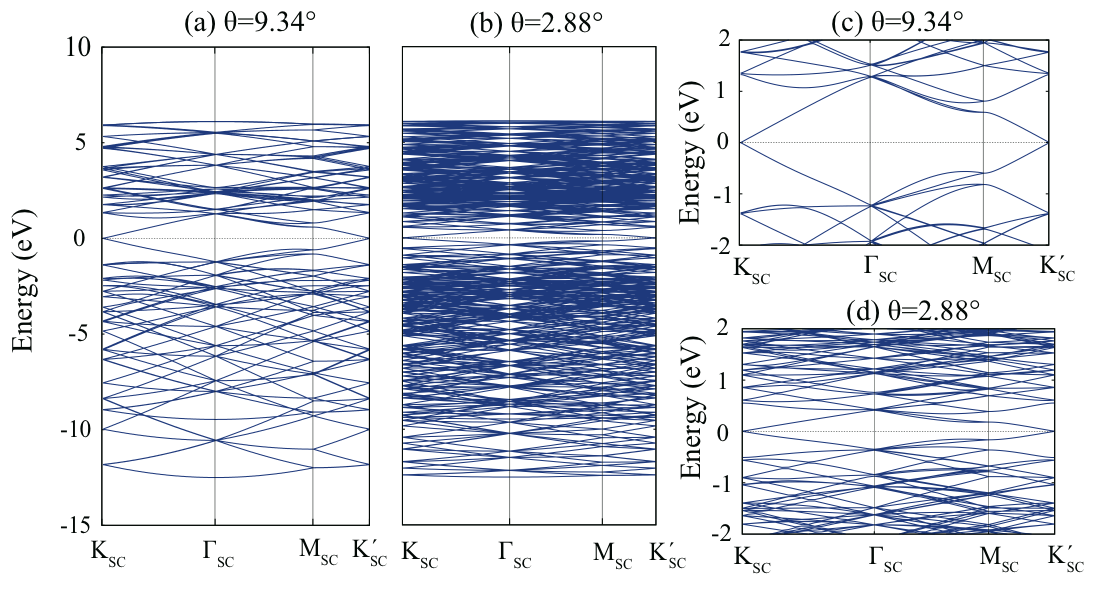}
\caption{(Color online) Calculated energy-band structures of tBLG with $\theta=9.34^{\circ}$ (a) and $\theta=2.88^{\circ}$ (b) by the TB calculation. Enlarged band structures near $E_{\rm F}$ are shown in (c) and (d).}
\label{fig:sc-band}
\end{figure}

\subsection{Unfolded bands of twisted bilayer graphene}

\begin{figure}
\includegraphics[bb = 0 0 372 276,width=0.8\linewidth]{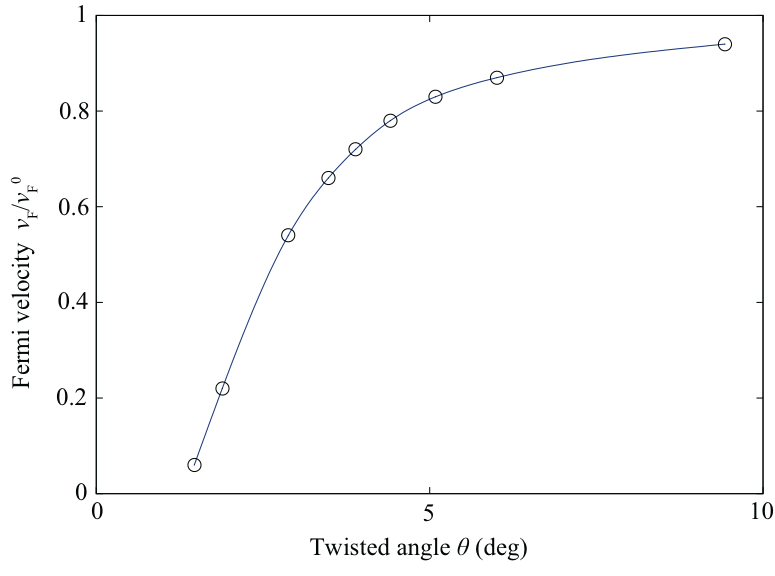}
\caption{(Color online) Fermi velocity of tBLG normalized to that of monolayer as a function of the twisted angle $\theta$.}
\label{fig:vf-angle}
\end{figure}

We start with the energy bands of tBLG represented in SBZ. Figure \ref{fig:sc-band} shows calculated energy bands of tBLG with the twisted angles of $\theta=9.34^{\circ}$ and $\theta=2.88^{\circ}$, along with their enlargements near $E_{\rm F}$. Looking at the vicinity of the Fermi level, the linear dispersion around $E_{\rm F}$ is observed also in tBLGs with these twisted angles. The Fermi velocity $v_{\rm F}$ is calculated from the gradient of the linear dispersion. The calculated $v_{\rm F}$ as a function of $\theta$ is shown in Fig. \ref{fig:vf-angle}: $v_{\rm F}$ remains 94\% at $\theta=9.34^{\circ}$ but decreases down to 6\% at $\theta=1.47^{\circ}$ of $v_{\rm F}$ of the monolayer graphene. The calculated values obtained here show fair agreement with the results in the past \cite{Exfoliation1,SiC3,Uchida,Neto2007,Pankratov2010,Magaud2010,Barticevic2010,MacDonald2011,Magaud2012}.

\begin{figure}
\includegraphics[bb= 0 0 416 343,width=1.0\linewidth]{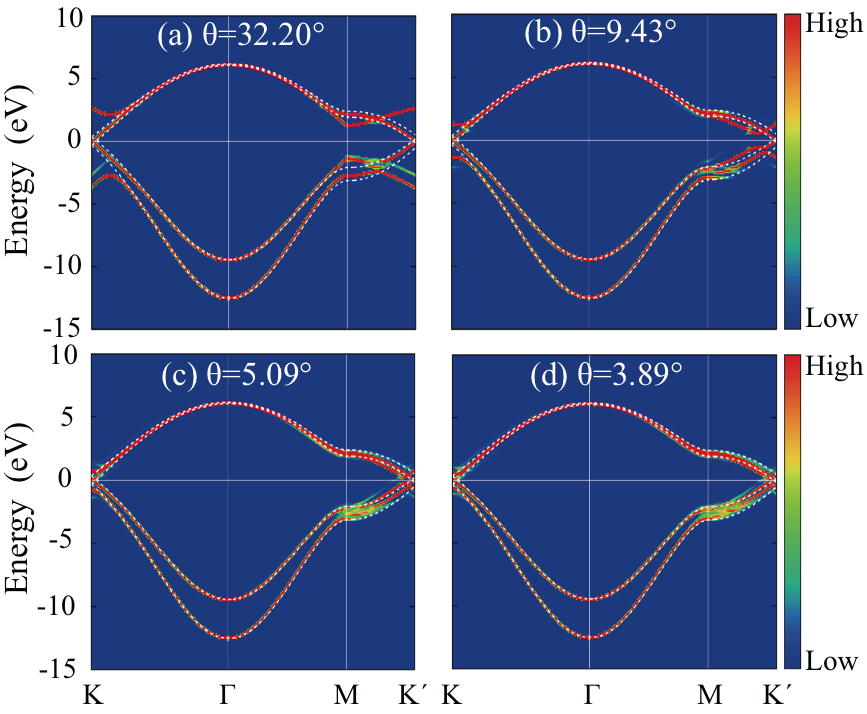}
\caption{(Color online) Contour plots of the unfolded bands, i.e.. the spectral function of tBLG with $\theta=32.20^{\circ}$ (a), $\theta=9.43^{\circ}$ (b), $\theta=5.09^{\circ}$ (c) and $\theta=3.89^{\circ}$ (d). The energy bands in the SBZ is unfolded to the PBZ of a monolayer graphene shown in red line in Fig. 1(b). The energy bands of the AA-stacking bilayer are shown by white dotted lines.}
\label{fig:ub-high}
\end{figure}

Figure \ref{fig:ub-high} shows the energy bands of several tBLGs unfolded to the red PBZ. We have found that the unfolded energy bands for these tBLGs are similar to each other especially in the region of the PBZ far from $K$ point. In this region of the PBZ, the spectral function is high only in the energy region of more than 2 eV above and below $E_{\rm F}$ (high-energy region). We also show the energy bands of the AA-stacking BLG as a reference in Fig. \ref{fig:ub-high}. We have found that the unfolded energy bands are almost identical to the energy band of the AA-stacking BLG in the high-energy region, irrespective of the twisted angle. It is of note that the energy bands of the AB-stacking BLG are also almost identical to those of the AA-stacking BLG in this energy region. The unfolded energy bands of tBLG obtained here reflect a fact that the energy spectrum of tBLG with the high energy region is determined primarily by the bonding and the antibonding characters of the in-plane $\pi$ orbitals. The way of stacking of two graphene layers is unimportant in the high-energy region.

\begin{figure}
\includegraphics[bb=0 0 488 414,width=1.0\linewidth]{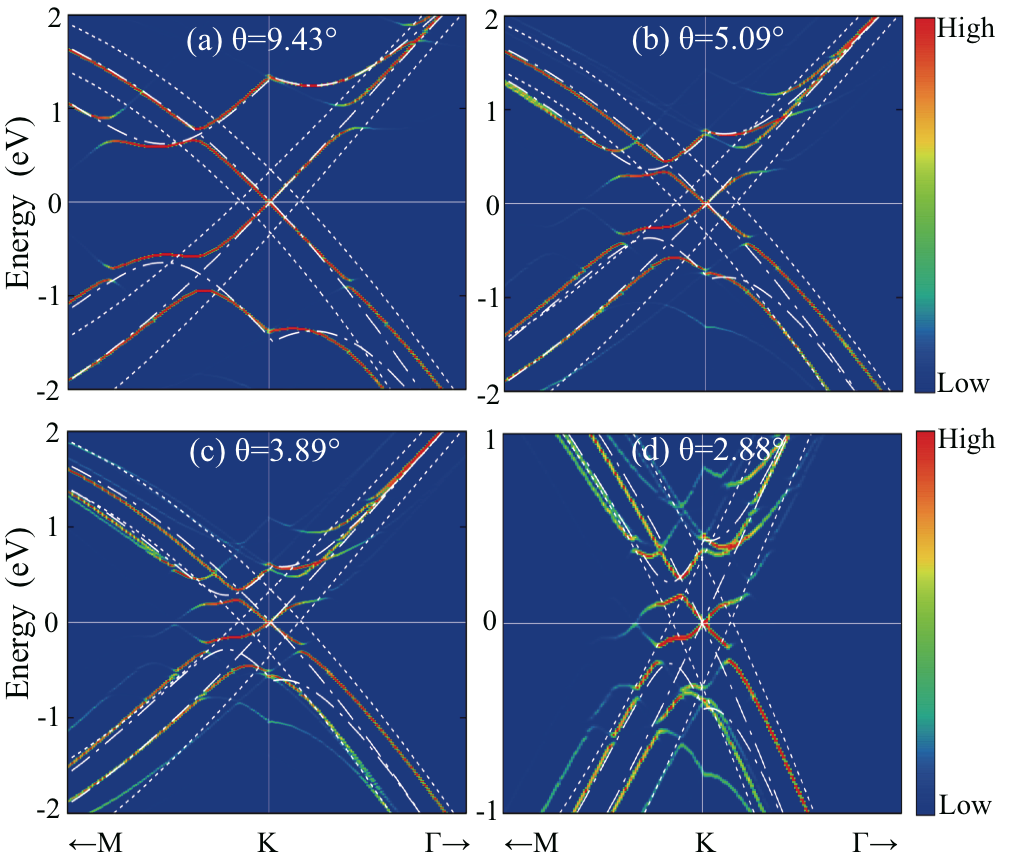}
\caption{(Color online) Contour plots of the unfolded bands of tBLGs near the Fermi energy with $\theta=9.43^{\circ}$ (a), $\theta=5.09^{\circ}$ (b), $\theta=3.89^{\circ}$ (c), and $\theta=2.88^{\circ}$ (d). The energy bands in the SBZ is unfolded to the PBZ of a monolayer graphene shown in red line in Fig. \ref{fig:atom-bz} (b). The energy bands of the tBLG having no interlayer interaction 
and of the AA-stacking bilayer are also shown by white dash-dotted lines and by white dotted lines, respectively.
}
\label{fig:ub-fermi}
\end{figure}

In contrast to the high-energy region above, the energy spectra near the Fermi level, i.e., within about 1 eV from $E_{\rm F}$, are drastically affected by the twisted angle in tBLG. Figure \ref{fig:ub-fermi} shows the obtained energy bands of tBLG unfolded in the vicinity of the $K$ point of the red PBZ. We have also shown the unfolded energy bands of tBLG having no interlayer interaction for comparison, along with the energy bands of the AA-stacking BLG. The energy bands with the linear dispersion are found around the $K$ point near $E_{\rm F}$. This is obviously the Dirac cone represented in the red PBZ: Namely the Dirac cone from one of the two graphene monolayers. We have found another Dirac cone with the hyperbolic dispersion near the $K$ point in the unfolded energy bands.
This Dirac cone comes from a Dirac cone of the other graphene monolayer, i.e., the Dirac cone located at the $K$ point of the blue PBZ. The $\Gamma$-$K$-$M$ line in the red PBZ is dislocated from the $\Gamma$-$K$-$M$ line in the blue PBZ, leading to the hyperbolic dispersion. The amount of the dislocation decreases with the decreasing $\theta$. This statement is corroborated by comparing the unfolded energy bands of tBLG with and without the interlayer interaction: In both cases, the linear and the non-linear bands emerge near $E_{\rm F}$.

\begin{figure}
\includegraphics[bb=0 0 439 195,width=1.0\linewidth]{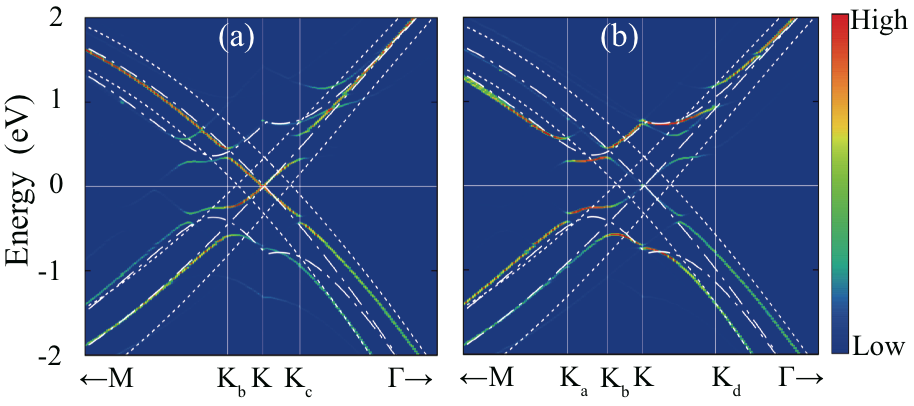}
\caption{(Color online) 
Contribution from each graphene monolayer to the spectral function of tBLG with $\theta=5.09^{\circ}$ unfolded to the red PBZ. The contribution from one monolayer corresponding to the red PBZ and the contribution from the other monolayer corresponding to the blue PBZ are shown in (a) and (b), respectively. The unfolded energy bands of the tBLG having no interlayer interaction and the energy bands of the AA-stacking BLG are shown by white dash-dotted and white dotted lines, respectively.} 
\label{fig:ub-layer}
\end{figure}

More importantly, we have found that the unfolded energy bands of tBLG exhibit the energy gap of about a half eV at certain $k$ points near the $K$ along the lines $K\Gamma$ and $KM$ (Fig.~\ref{fig:ub-fermi}). The position of the gap decreases from more than 1 eV for the $\theta= 9.43^{\circ}$ to less than half eV for the $\theta= 2.88^{\circ}$. The energy gap in the spectral function (the unfolded band) means that the particular states labelled by $\kvec$ in the PBZ are strongly modified by the super-periodicity, i.e., the Moir\`e  pattern. We have also found that the $k$ point at which the gap opens comes close to the $K$ when $\theta$ decreases. The gap opening inevitably makes the dispersion flat in the vicinity, and thus reduces the Fermi-level velocity as this gap-opening $k$ point comes close to the $K$ point.

It is very informative to decompose the spectral function (the unfolded band) into two components, i.e., the contribution from each graphene monolayer in tBLG. This decomposition is possible without any ambiguity because the summation in Eq. (\ref{eq:TB}) runs over atom indexes belonging to each monolayer. Figure \ref{fig:ub-layer} shows such decomposition of the spectral function (the unfolded band) of tBLG with the twisted angle of $\theta=5.09^{\circ}$ unfolded to the red PBZ. As clearly seen, the spectral weight of the Dirac cone at $K$ point comes from one of the monolayer corresponding to the red PBZ, while the energy bands with the hyperbolic dispersion is attributed to the Dirac cone from the other monolayer corresponding to the blue PBZ as mentioned in the previous paragraph. Interestingly, we have found that the $k$ positions at which the energy gap opens exhibit strong layer dependence: The energy gap opens at $K_b$ and $K_c$ points in the contribution from the layer corresponding to the red PBZ, whereas it does at $K_a$, $K_b$ and $K_d$ points in the contribution from the layer corresponding to the blue PBZ. 
The origin of the gap opening and this layer dependence will be elucidated below.

\subsection{Reciprocal-space resolved Dirac-cone interaction}
\label{sec:r-space}

\begin{figure*}
\centering
\includegraphics[bb=0 0 660 455,width=0.9\linewidth]{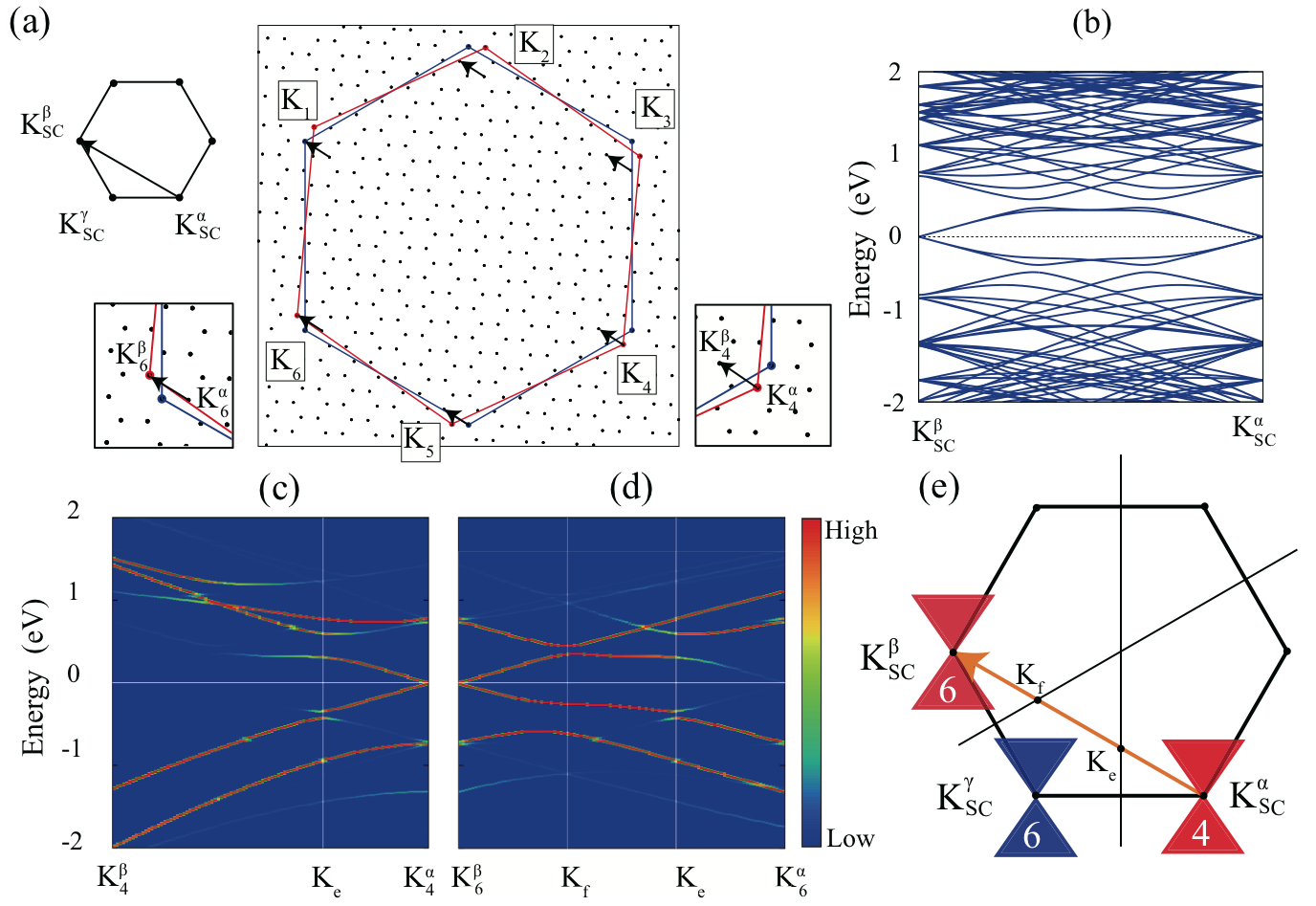}
\caption{(Color online) 
(a) Arrangements of the supercell Brillouin zone (SBZ) and the primitive BZs (PBZs) with several $K$ and $K^{\prime}$ points. Black dots depict $K_{\rm SC}$ and $K^{\prime}_{\rm SC}$ points at the corners of SBZs and the red and blue PBZs twisted to each other are shown by red and blue lines, respectively. The $K$ and $K^{\prime}$ points of the PBZs are labeled as $K_l$ ( $l$ = 1 -- 6 ) as in the central figure. The enlargements near the $K_4$ and the $K_6$ are also shown. The symmetry lines discussed in the text are shown by black arrows. (b) Energy bands along the symmetry line depicted by the black arrow in SBZ. (c) and (d): The spectral functions along the same symmetry lines which are unfolded near the region of the $K_4$ point (c) and the region of the $K_6$ point (d). (e) Dirac cones at $K_4$ and $K_6$ points in red and blue PBZs folded in the SBZ. The $K_e$ and $K_f$ points lie on the perpendicular bisector of each two Dirac cones.
} 
\label{fig:layer-split}
\end{figure*}

We are now in a position to clarify the microscopic reason for the gap opening described in the preceding subsection and the consequent reduction of the Fermi velocity in tBLG with small twisted angle. Suppose one $K_{\rm SC}$ point labelled as $K^{\alpha}_{\rm SC}$ and another labelled as $K^{\beta}_{\rm SC}$ in SBZ and the line connecting these two $K_{\rm SC}$ points [Fig. \ref{fig:layer-split}(a)]. We also label adjacent $K_{\rm SC}$ point as $K^{\gamma}_{\rm SC}$. When we unfold this line to the vicinity of the three $K$ points and three $K^{\prime}$ points in the PBZs [labelled as $K_l(l=1-6)$ in Fig. \ref{fig:layer-split}(a) ], the generated six lines are inequivalent to each other in the PBZs.

Figure \ref{fig:layer-split}(b) shows the calculated energy bands along the $K^{\alpha}_{\rm SC}$- $K^{\beta}_{\rm SC}$ line in the SBZ. The Dirac cone folded from the PBZ is observed at both $K^{\alpha}_{\rm SC}$ and $K^{\beta}_{\rm SC}$ points. We now unfold these energy bands to the vicinity of the $K_4$ point [Fig. \ref{fig:layer-split}(c)] and the $K_6$ point [Fig. \ref{fig:layer-split}(d)] in the PBZ. We have found that the Dirac cone around $K^{\alpha}_{\rm SC}$ has the spectral weight solely from the $K_4$ point in the red PBZ (labelled as $K^{\alpha}_4$): The vanishing spectrum weight from the $K_6$ point is shown in Fig. \ref{fig:layer-split}(d) (The unfolded energy bands around other $K_l$ points are not shown). This is a consequence that other point unfolded from $K^{\alpha}_{\rm SC}$ to the vicinity of $K_l$ ( $l$ = 1, 2, 3, 5 or 6) point is dislodged from the $K_l$ point of the red PBZ [Fig.\ref{fig:layer-split}(a)] and thus corresponding energy is far from $E_{\rm F}$. It is noteworthy that the $K_5$ point in the blue PBZ has the spectral weight to the Dirac cone around $K^{\alpha}_{\rm SC}$. 
Moving from $K^{\alpha}_{\rm SC}$ to $K^{\beta}_{\rm SC}$, the spectral weight in the low energy region (about 0.3 eV from $E_{\rm F}$) becomes small in the vicinity of $K_4$ point and in turn increases in the vicinity of the $K_6$ point. [Figs. \ref{fig:layer-split}(c) and(d)]. 
This is a consequence of a fact that the Dirac cones at the $K_6$ points in red PBZ and the blue PBZ are folded on the $K^{\beta}_{\rm SC}$ and the $K^{\gamma}_{\rm SC}$ points, respectively and thus contribute to the low-energy spectrum on the $K^{\alpha}_{\rm SC}$- $K^{\beta}_{\rm SC}$ line.  
Finally around $K^{\beta}_{\rm SC}$ point, the Dirac cone comes solely from $K^{\beta}_6$ point in red PBZ (and $K^{\beta}_1$ point in the blue PBZ).

The unfolded energy bands show the energy gap around 0.3 eV below and above $E_{\rm F}$ at the point $K_e$ in the vicinity of the $K_4$ point [Fig. \ref{fig:layer-split}(c)] and at the points $K_e$ and $K_f$ in the vicinity of the $K_6$ point [Fig. \ref{fig:layer-split}(d)]. 
We have found that the points $K_e$ and $K_f$ are the intersections of the line $K^{\alpha}_{\rm SC}$ - $K^{\beta}_{\rm SC}$ and the perpendicular bisectors of sides of the hexagonal SBZ, as shown in Fig. \ref{fig:layer-split}(e). 
By examining relative arrangements of SBZs and PBZs shown in Fig. \ref{fig:layer-split}(a), it becomes clear that the Dirac cone located at $K^{\gamma}_{\rm SC}$ has the spectral weight from the Dirac cones at $K_5$ in the red PBZ and at $K_6$ in the blue BZ, as is explained above for the cases of $K^{\alpha}_{\rm SC}$ and $K^{\beta}_{\rm SC}$. Therefore, along the line $K^{\alpha}_{\rm SC}$ - $K^{\beta}_{\rm SC}$, the Dirac cones folded to $K^{\alpha}_{\rm SC}$ and those folded to $K^{\gamma}_{\rm SC}$ interact to each other strongly at the point $K_e$ since all the four linear bands has the same energy at $K_e$. This is the reason for the gap opening. Similarly, the interactions among the Dirac cones folded on $K^{\beta}_{\rm SC}$ and on $K^{\gamma}_{\rm SC}$ causes the gap opening at the $K_f$ point.

\begin{figure}
\includegraphics[bb=0 0 371 270,width=0.8\linewidth]{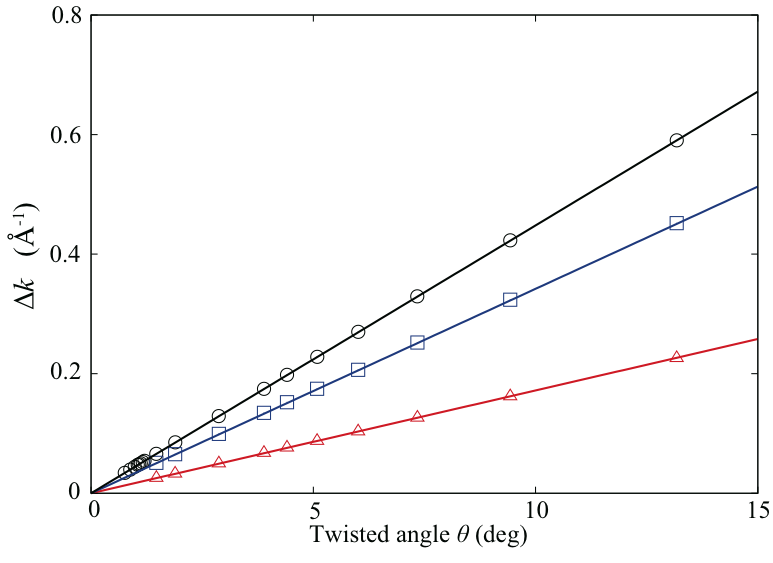}
\caption{(Color online) Variation of the energy gap points $K_e$ (blue square) and $K_f$ (red triangle) as a funciton of the twisted angle in tBLG. The plotted are the distances $\Delta k$ of those points from the symmetry point $K^{\alpha}_{\rm SC}$ (see text). The side lengths of the supercell BZ (SBZ) are also shown by black circles.}
\label{fig:split-size}
\end{figure}

Figure \ref{fig:split-size} shows the twisted-angle dependence of the $K_e$ and $K_f$ positions.
We plot the distance between $K_e$ or $K_f$ and $K_{\rm SC}^{\alpha}$, along with the side length of SBZ. As is clearly shown, the size of the SBZ and the distances for $K_e$ and $K_f$ exhibit linear relation with respect to the twisted angle. This behavior clearly means that positions at which the energy gaps open are unchanged relative to SBZ size. This result strongly corroborates our statement that the gap opening takes place at the intersections with the perpendicular bisectors, thus being caused by the reciprocal-space interactions of Dirac cones.

\begin{figure}
\includegraphics[bb=0 0 221 233,width=0.7\linewidth]{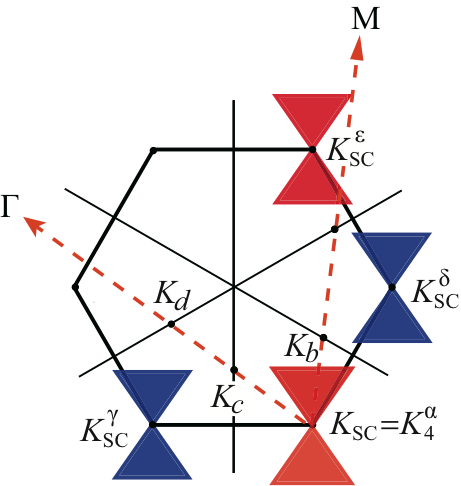}
\caption{(Color online) 
Supercell Brillouin zone (SBZ) near the $K_4$ point in Fig.~\ref{fig:layer-split} (a). The symmetry lines $KM$ and $K\Gamma$ along which the unfolded bands are calculated in Fig.~\ref{fig:ub-layer} are shown by (red) dashed lines. 
} 
\label{exp:ub-layer}
\end{figure}

This reciprocal-space resolved interaction between Dirac cones also elucidates the layer dependence of the gap-opening position shown in Fig.~\ref{fig:ub-layer} of the preceding subsection. Figure \ref{exp:ub-layer} shows the SBZ in the unfolded region near the $K_4$ point in the PBZ. The symmetry lines $KM$ and $K\Gamma$ along which the unfolded bands in Fig.~\ref{fig:ub-layer} are calculated are also shown. From Fig.~\ref{fig:layer-split} (a), it is clear that the Dirac cone at $K^{\alpha}_{\rm SC}$ is folded from the cones at $K_4$ in the red PBZ and $K_5$ in the blue PBZ. Similarly, the cones at $K^{\gamma}_{\rm SC}$ are from the cones at $K_6$ in the blue PBZ and $K_5$ in the red PBZ. Other $K$ points in Fig.~\ref{exp:ub-layer}, i.e., $K^{\delta}_{\rm SC}$ and $K^{\epsilon}_{\rm SC}$, are from those at $K_4$ in the blue PBZ and $K_3$ in the red PBZ and from those at $K_2$ in the red PBZ and $K_3$ in the blue PBZ, respectively. 
\par
We first focus the contribution from the monolayer corresponding to the red PBZ [Fig.~\ref{fig:ub-layer}(a)]. At the $K$ point (i.e., $K_4^{\alpha}$) the unfolded bands exhibit a Dirac cone at $E_{\rm F}$ originating from the cone at $K_4$ in the red PBZ. Moving from $K$ to $M$, this cone interacts with another cone at $K_{\rm SC}^{\delta}$ (i.e., the cone at $K_4$ in blue PBZ) and opens a gap at the $K_b$ point which is the intersection of the $KM$ line and the perpendicular bisector. Moving further from $K_e$ point, the spectral weight is shifted to the higher energy region and no appreciable cone-cone interaction exist. Along the $K\Gamma$ line, the cone at $K_4^{\alpha}$ start to interact with the cone at $K_{\rm SC}^{\gamma}$ and opens a gap at the intersection $K_c$. Moving further to $\Gamma$, the spectral weight in the low-energy region diminishes and thus no appreciable cone-cone interaction exist. 
\par
We next consider the contribution from the monolayer corresponding to the blue PBZ [Fig.~\ref{fig:ub-layer}(b)]. At the $K_4^{\alpha}$ point the unfolded bands show a gap. This is a gap between the upper and lower Dirac cone located at the $K_{\rm SC}^{\delta}$ (i.e. the $K_4$ in the blue PBZ). Moving from $K$ to $M$, this Dirac cone interacts with the cone at $K_4^{\alpha}$ and opens a gap at the intersection $K_b$. Moving further from $K_b$, the cone at $K_{\rm SC}^{\delta}$ then interacts with the cone at $K^{\epsilon}_{\rm SC}$ and opens a gap at the intersection $K_a$. Along the $K\Gamma$ line, the Dirac cone at the $K_4$ in the blue PBZ produces the spectral weight in rather higher energy region and no appreciable interaction with the cones near the $K\Gamma$ line exist. However, in the higher energy region around 1 eV, this cone interacts with the corn located at $K^{\epsilon}_{\rm SC}$ and opens a gap at $K_d$ which is the intersection of the $K\Gamma$ line and the perpendicular bisector.

\begin{table}
\caption{Calculated energy gaps of tBLGs with several twisted angles $\theta$ in the valence-band region (subscript $v$) with using the interlayer interaction (subscript inter) and the total Hamiltonian at the $K_e$ and the $K_f$ points in unit of eV.} 
\label{table1}
  \begin{ruledtabular}
  \begin{tabular}{c|cccc|cccc}
  \hline
   &\multicolumn{4}{c|}{$K_e$} & \multicolumn{4}{c}{$K_f$} \\
  $\theta$&$\Delta E_v$&$\Delta E_v^{\rm inter}$&$\Delta E_c$&$\Delta E_c^{\rm inter}$&$\Delta E_v$&$\Delta E_v^{\rm inter}$&$\Delta E_c$&$\Delta E_c^{\rm inter}$ \\  
  \hline
  $2.88^{\circ}$ & 0.080 & 0.059 & 0.270 & 0.122 & 0.314 & 0.164 & 0.094 & 0.077 \\ 
  $3.89^{\circ}$ & 0.084 & 0.059 & 0.289 & 0.247 & 0.334 & 0.246 & 0.100 & 0.087 \\ 
  $5.09^{\circ}$ & 0.088 & 0.063 & 0.290 & 0.269 & 0.358 & 0.312 & 0.105 & 0.096 \\ 
  $9.43^{\circ}$ & 0.094 & 0.067 & 0.273 & 0.266 & 0.414 & 0.395 & 0.110 & 0.106 \\ 
  \hline
  \end{tabular}
  \end{ruledtabular}
\end{table}

Another interesting feature in the unfolded energy bands [Figs. \ref{fig:layer-split}(c) and (d)] is that the obtained energy gaps are different between the valence and the conduction bands To elucidate this feature, we decompose the Hamiltonian of tBLG into the diagonal and off-diagonal parts;
\begin{eqnarray*}
   H &=& \begin{pmatrix}
        H_{U} & H_{LU}  \\
        H_{UL} & H_{L} 
       \end{pmatrix} \\ 
     &=& H_0 + V  
      =\begin{pmatrix}
        H_{U} & 0 \\
        0 && H_{L}
        \end{pmatrix}
     +  \begin{pmatrix}
        0 & H_{LU}  \\
        H_{UL} & 0
       \end{pmatrix}.
\end{eqnarray*}

where $H_U$ is the Hamiltonian of one of the two graphene monolayer (the upper layer Hamiltonian), $H_L$ the Hamiltonian of the other layer (the lower layer Hamiltonian), and $H_{UL}$ and $H_{LU}$ interlayer interaction. This decomposition is carried out unambiguously in the tight binding model. The interlayer-interaction energy of $I$-th band, $E^{\rm inter}_I$, is defined as $E^{\rm inter}_I=\braket{\Psi_I|V|\Psi_I}$ using the eigenstate of the Hamiltonian, $\Psi_I$. The total Hamiltonian energy of $I$-th band $E_I$ is also defined as $E_I=\braket{\Psi_I|H|\Psi_I}$. Then we take the upper band and the lower band at the energy gap and calculate the difference in the interlayer interaction energy $\Delta E^{\rm inter}$ and the total Hamiltonian energy $\Delta E$. Then $\Delta E$ corresponds to the energy gap in the unfolded energy bands. We calculate those values for the conduction-band region, $\Delta E^{\rm inter}_c$ and $\Delta E_c$, and for the valence-band region, $\Delta E^{\rm inter}_v$ and $\Delta E_v$, at the $K_e$ and $K_f$ points. Table \ref{table1} shows calculated $\Delta E^{\rm inter}_c$, $\Delta E_c$, $\Delta E^{\rm inter}_v$ and $\Delta E_v$ of tBLGs for several twisted angles $\theta$. We have found that the twisted-angle dependence of each quantity is relatively small. We have also found that the contribution from the interlayer interaction to the energy gap is substantially large about more than 70$\%$. Interaction between the Dirac cones from adjacent layers is important for understanding of the energy gap obtained in the present calculation.

\section{Conclusion}\label{sum}

We have calculated the energy spectrum of electrons in twisted bilayer graphene (tBLG) by the band-unfolding method in the tight-binding model and clarified the drastic modulation of the electronic structure, in particular the band-gap opening at special $k$ points and the resulting electron-velocity reduction at the Fermi level, in the tBLG with small twisted angles. The tBLGs with small twisted angles are accompanied with Moir\`e pattern which produces particular super-periodicities. Our calculated unfolded bands, i.e., the spectral function in terms of the Bloch functions of the primitive Brillouin zone in each graphene monolayer, clarifies that Dirac cones of each monolayer twisted to each other interact at some particular points in the reciprocal space due to the Moir\`e-induced super-periodicity and thus opens a gap at the points. We have clarified that the particular points where the cone-cone interaction becomes important is determined by the geometrical arrangements of the two graphene monolayers twisted to each other. We have unequivocally shown that the gap-opening points become close to the $K$ points with decreasing the twisted angle and thus the Fermi-level velocity decreases drastically with the decreasing the twisted angle. When the two layers are not twisted to each other, this cone-cone interaction disappears suddenly and the drastic modulation of the electronic structure also disappears. 

\begin{acknowledgments}
This work was supported in part by MEXT as a social and scientific priority issue (Creation of new functional devices and high-performance materials to support next-generation industries) to be tackled by using post-K computer. Computations were performed mainly at the Supercomputer Center at the Institute for Solid State Physics, The University of Tokyo, The Research Center for Computational Science, National Institutes of Natural Sciences, and the Center for Computational Science, University of Tsukuba.
\end{acknowledgments}


\begin{thebibliography}{99}

\bibitem{SW}
J. C. Slonczewski and P. R. Weiss, Phys. Rev. {\bf 109}, 272, (1958).


\bibitem{QHE1} 
K. S. Novoselov, A. K. Geim, S. V. Morozov, D. Jiang, M. I. Katsnelson, I. V. Grigorieva, S. V. Dubonos and A. A. Firsov, Nature {\bf 438}, 197 (2005).

\bibitem{QHE2} 
Y. Zhang, Y.-W. Tan, H. L. Stormer and P. Kim, Nature {\bf 438}, 201 (2005).

\bibitem{louie}
Y.-W. Son, M. L. Cohen, and S. G. Louie, Nature, {\bf 444}, 347 (2007).

\bibitem{okada}
S. Okada and A. Oshiyama, Phys. Rev. Lett. {\bf 87}, 146803 (2001).

\bibitem{device}
S. Bae, H. Kim, Y. Lee, X. Xu, J.-S. Park, Y. Zheng, J. Balakrishnan, T. Lei, H. R. Kim, Y. I. Song, Y.-J. Kim, K. S. Kim, B. Ozyilmaz, J.-H. Ahn, B. H. Hong and S. Iijima, Nature Nanotechnology, {\bf 5}, 574 (2010).

\bibitem{Exfoliation1} 
Z. Ni, Y. Wang, T. Yu, Y. You, and Z. Shen, Phys. Rev. B {\bf 77}, 235403 (2008).

\bibitem{Exfoliation2}
P. Poncharal, A. Ayari, T. Michel, and J.-L. Sauvajol, Phys. Rev. B {\bf 78}, 113407 (2008).

\bibitem{SiC1}
J. Hass, F. Varchon, J. E. Millan-Otoya, M. Sprinkle, N. Sharma, W. A. de Heer, C. Berger, P. N. First, L. Magaud, and E. H. Conrad, Phys. Rev. Lett. {\bf 100}, 125504 (2008).

\bibitem{SiC2}
F. Varchon, P. Mallet, L. Magaud, and J.-Y. Veuillen, Phys. Rev. B {\bf 77}, 165415 (2008).

\bibitem{SiC3}
A. Luican, Guohong Li, A. Reina, J. Kong, R. R. Nair, K. S. Novoselov, A. K. Geim, and E. Y. Andrei, Phys. Rev. Lett. {\bf 106}, 126802 (2011).

\bibitem{ABband}
S. Latil and L. Henrard, Phys. Rev. Lett. {\bf 97}, 036803 (2006).

\bibitem{STM2005} 
W. T. Pong and C. Durkan, J. Phys. D {\bf 38}, R329 (2005).

\bibitem{STM1993} 
Z. Y. Rong and P. Kuiper, Phys. Rev. B {\bf 48}, 17427 (1993).

\bibitem{ARPES2012}
T. Ohta, J. T. Robinson, P. J. Feibelman, A. Bostwick, E. Rotenberg, and T. E. Beechem, Phys. Rev. Lett. {\bf 109}, 186807 (2012).

\bibitem{hicks}
J.~Hicks, M.~Sprinkle, K.~Shepperd, F.~Wang, A.~Tejeda, A.~Taleb-Ibrahimi, F.~Bertran, P.~Le F\'evre, W.~A.~de Heer, C.~Berger, E.~H.~Conrad, Phys. Rev. B {\bf 83}, 205403 (2011).

\bibitem{Uchida} 
K. Uchida, S. Furuya, J.-I. Iwata, and A. Oshiyama, Phys. Rev. B {\bf 90}, 155451 (2014).

\bibitem{ARPES2009}
M. Sprinkle, D. Siegel, Y. Hu, J. Hicks, A. Tejeda, A. Taleb-Ibrahimi, P. Le Fevre, F. Bertran, S. Vizzini, H. Enriquez, S. Chiang, P. Soukiassian, C. Berger, W. A. de Heer, A. Lanzara, and E. H. Conrad, Phys. Rev. Lett. {\bf 103}, 226803 (2009).

\bibitem{latil}
S. Latil, V. Meunier, and L. Henrard, Phys. Rev. B {\bf 76}, 201402(R) (2007).

\bibitem{Pankratov2008} 
S. Shallcross, S. Sharma, and O. A. Pankratov, Phys. Rev. Lett. {\bf 101}, 056803 (2008).

\bibitem{Neto2007} 
J. M. B. Lopes dos Santos, N. M. R. Peres, and A. H. Castro Neto, Phys. Rev. Lett. {\bf 99}, 256802 (2007).

\bibitem{Pankratov2010} 
S. Shallcross, S. Sharma, E. Kandelaki, and O. A. Pankratov, Phys. Rev. B {\bf 81}, 165105 (2010).

\bibitem{Magaud2010} 
G. Trambly de Laissardiere, D. Mayou, and L. Magaud, Nano Lett. {\bf 10}, 804 (2010).

\bibitem{Barticevic2010} 
E. Suarez Morell, J. D. Correa, P. Vargas, M. Pacheco, and Z. Barticevic, Phys. Rev. B {\bf 82}, 121407(R) (2010).

\bibitem{MacDonald2011} 
R. Bistritzer and A. H. MacDonald, Proc. Nat. Acad. Sci. {\bf 108}, 12237 (2011).

\bibitem{Magaud2012} 
G. Trambly de Laissardiere, D. Mayou, and L. Magaud, Phys. Rev. B {\bf 86}, 125413 (2012).

\bibitem{Neto2012} 
J. M. B. Lopes dos Santos, N. M. R. Peres, and A. H. Castro Neto, Phys. Rev. B {\bf 86}, 155449 (2012).

D





\bibitem{computics}
{\it Materials Design through Computics}, Scientific Research on Innovation Areas, MEXT Grant-in-Aid Project, Japan (http://computics-material.jp/index-e.html). 

\bibitem{hasegawa}
Y. Hasegawa, J.-I. Iwata, M. Tsuji, D. Takahashi, A. Oshiyama, K. Minami, T. Boku, H. Inoue, Y. Kitazawa, I. Miyoshi, M. Yokokawa, Int. J. High Performance Computing Applications, {\bf 28}, 335 (2014). 

\bibitem{Ku2010} 
W. Ku, T. Berlijn and C.-C. Lee, Phys. Rev. Lett. {\bf 104} 216401 (2010).

\bibitem{Zunger2010} 
V. Popescu and A. Zunger, Phys. Rev. Lett. {\bf 104}, 236403 (2010).

\bibitem{Zunger2012} 
V. Popescu and A. Zunger, Phys. Rev. B {\bf 85}, 085201 (2012).

\bibitem{Sawatzkey} 
M. W. Haverkort, I. S. Elfimov, and G. A. Sawatzky, arXiv:1109.4036 (2011).

\bibitem{Ozaki} 
C.-C. Lee, Y. Yamada-Takamura and T. Ozaki, J. Phys. Condens. Matter, {\bf 25} 345501 (2013).

\end{thebibliography}
\end{document}